\documentstyle[epsf]{l-aa}

\begin{document}

   \thesaurus{06         
              (2.12.1;  
               8.14.1;  
               13.25.5;  
               2.18.7 )}  

   \title{Cylotron line models for the X-ray pulsar A0535+26}

   \author{R. A. Araya\inst{1} \and A. K. Harding\inst{2}}

   \offprints{A. K. Harding}

   \institute{
              Department of Physics and Astronomy,
              Johns Hopkins University,
              Baltimore, MD 21218
              \and
              Laboratory for High Energy Astrophysics,
              NASA/Goddard Space Flight Center,
              Greenbelt, MD 20771
             }

   \date{Received ; accepted }

   \maketitle

\begin{abstract}

 The spectrum of the transient X-ray binary pulsar A0535+26 
 obtained by OSSE in February 1994 shows an absorption feature 
 at 110 keV but does not confirm a feature at around 55 keV, as
 previously reported by other instruments.  If the 110 keV 
 feature is due to cyclotron scattering at the first    
 harmonic, then the magnetic field required is about 
 $10^{13}$ Gauss, the highest observed in an X-ray pulsar.
 Conversely, if this strong feature is a second harmonic and the line
 formation process is such that an extremely weak fundamental results 
 at $\simeq$ 55 keV, the estimate of the field strength is halved.
 We present results of detailed cyclotron line transfer models
 from two source geometries to explore the theoretical contraints
 on the line shapes in this source.
 It is found that while a fundamental harmonic line at
 $\sim$ 55 keV may be partially filled-in by angle redistribution
 in cylindrical geometries, the required viewing angles give a 
 second harmonic line shape inconsistent with the observation.
 Interpretation of the feature at 110 keV as a first harmonic seen
 at small angles to the field yields more consistent line shapes.

   \end{abstract}
 
%
 
\section{Introduction}
 
 Cylotron line features have been detected in a number of X-ray pulsar
 spectra (Nagase 1989) and possibly also in the spectra of several 
 gamma-ray bursts (Murakami et al. 1988).  The line energies fall in the
 range from about 4 to 40 keV, indicating scattering of electrons in
 magnetic fields up to $3.5 \times 10^{12}$ G.  Detection of cyclotron
 scattering lines in spectra of astrophysical sources is extremely
 important because it is at present the only $direct$ method for
 measuring the magnetic field of a neutron star.
 
 The spectrum of the transient X-ray binary pulsar A0535+26 appears to
 show cyclotron features at the highest energies yet measured from any
 source.  The HEXE/TTM instrument on Mir/Kvant fit the spectrum observed
 from a 1989 outburst with two harmonic features at around 50 and 100 keV
 (Kendziorra et al. 1994).  While the 100 keV feature reached 5-6$\sigma$
 significance in some pulse phase intervals, the 50 keV feature was of
 very low ($< 3\sigma$) significance.  The OSSE instrument on CGRO
 observed a later outburst in 1994 and detected a highly significant 
 ($P < 10^{-15}$) feature in the phase averaged spectrum at 110 keV   
 (no phase resolved spectra are yet available), but no feature at 50-55 keV  
 (Grove et al. 1995).  The present observations alone cannot 
 determine whether the magnetic field in the source region is $B \sim
 5 \times 10^{12}$ G (5 TG), where the first cyclotron harmonic occurs
 at 55 keV,
 or $B \sim 10^{13}$ G (10 TG), where the first harmonic is at 110 keV.
 This situation raises an important theoretical question: is it
 possible for some combination of physical parameters and source geometry
 to produce a strong second harmonic and a very weak or {\em undetectable}
 first harmonic? 

     In the present work, the parameter space of a relativistic resonant
 scattering simulation is sampled to model the conditions leading 
 to the formation of
 the spectrum in A0535+26, for both low-field ($B \simeq 5$ TG) and high-field 
 ($B \simeq 11$ TG) cases. 
 In the low-field cases, the sensitivity of the line features to viewing
 angle with respect to the field allows a partial filling of the 
 fundamental harmonic by photon redistribution.  
 This effect along with the possible contribution of spawned photons
 from higher Landau levels are studied in detail.  In the high-field case,
 there is no harmonic feature at 55 keV and the 110 keV absorption 
 line is direct evidence for a magnetic field
 near the quantum critical field $B_{cr}=m^2c^3/e\hbar = 44.13$ TG.

     Resonant cyclotron features result from scattering of an incident
 photon spectrum by electrons in a
 magnetized plasma.  The discreteness of electronic energy levels in
 a uniform magnetic field produces
 harmonic features in the emergent spectra which are highly sensitive
 to viewing angle with respect to the field and plasma geometry.
     As the ratio ($B/B_{\rm cr}$) nears 1, the study of strongly magnetized
 systems motivates a relativistic approach to the relevant
 processes for the formation of cyclotron lines. 
 To model the spectra of A0535+26, we use a Monte Carlo resonant 
 transfer code which includes relativistic Quantum Electrodynamic magnetic
 cross sections allowing for the inclusion of up to four harmonics, 
 photon spawning and angle redistribution of photons. In the following
 section we give a brief description of the code; a detailed account
 will be given elsewhere.  Then, a report on trial runs to assess  
 the likelihood of a low field {\it versus} a high field scenario 
 is presented. More detailed fits, along with quantitative         
 estimates for the goodness of the fits, are 
 reported in Araya \& Harding (1996).
   

\section{Monte Carlo Resonant Transfer Code}

 The Monte Carlo code allows for the injection of individual
 photons and their diffusion in 3-D optical depth space. 
 Relativistic treatment of cyclotron scattering
 is required when the magnetic field involved becomes comparable with 
 $B_{\rm cr}.$  Our code uses the cyclotron scattering cross section
 calculated as a 2nd order QED process. 
 Natural line widths of the cross sections are introduced via radiative
 corrections to the electron propagator, using  
 solutions to the Dirac-Landau equation which diagonalize the 
 self-interaction (Herold, Ruder \& Wunner 1982).
 The corrections to the propagator introduce small imaginary terms in the
 resonant denominators, the lowest order being the cyclotron decay 
 rate (Graziani, Harding \& Sina 1995), resulting in a quasi-Lorentzian 
 line shape.
 This prescription is used in our code for the scattering mean-free paths.
 A Lorentzian line width (absorption approximation: Harding \& Daugherty 
 1991) is used to obtain
 the momentum of the scattering electron from the line profiles. 

    Furthermore, the code assumes a relativistic Maxwellian electron
 distribution function and includes excitation of higher Landau
 levels and the resultant photon spawning through cyclotron decay
 of these levels. Also, an approximate relativistic scattering angle
 redistribution of photons is used in place of the full differential
 cross section.

   \begin{figure*}
\epsfysize=12cm 
\epsfbox{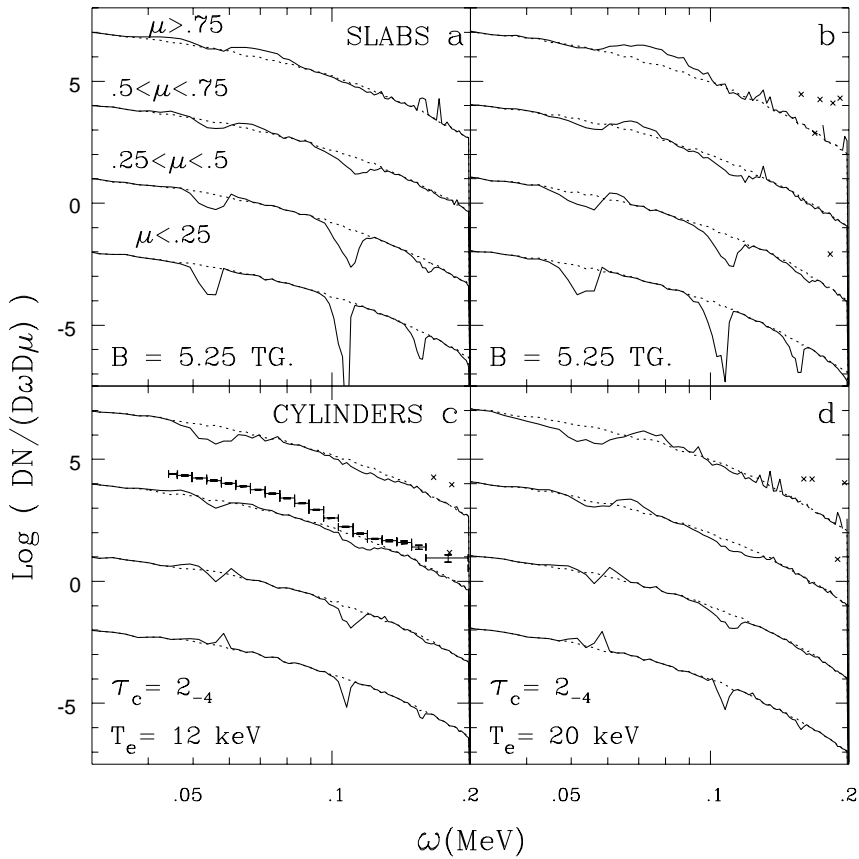}   
\vskip -12cm
\hfill      \parbox[b]{5cm}{\caption{
 Angle dependent model photon spectra for the lower field runs.
   Each run results from 50 to 100 thousand photons injected isotropicaly.
   $\mu = {\rm cos}\theta$. {\bf Dotted line}: injected continuum. 
   {\bf Solid line}: output scattered spectrum. {\bf Crosses}: 
   single upscattered
   photons. {\bf Horizontal error bars}: OSSE A0535+26 spectrum.
   Spectra from slabs are on the top and from cylinders on the 
   bottom. Columns share the values of $\tau_c,~T_e,~T_c,~\alpha~{\rm and}~B$.
   $T_c=14.5$ keV and $\alpha = 0$ for all four runs. 
   See text for notation. The flux normalization is arbitrary.
   These are local spectra emitted near the neutron star
 surface and do not include General relativistic effects such as red shift
 and light bending.  
   }}
            
\vskip 3.5cm   
         \label{Fig1}%
    \end{figure*}

\section{Model Calculations}

    We consider two geometries for the scattering region: plane slab, with
 the magnetic field parallel to the slab normal, and cylindrical, with
 the magnetic field parallel to the cylinder axis.  An isotropic
 continuum photon spectrum is incident from a source at the slab midplane
 or along the cylinder axis. The escaping photons are accumulated in
 four ranges of $\mu$, the cosine of the viewing 
 angle to the field. Following Grove et al. (1994), the
 continuum spectrum is modeled as a power law times an exponential:
 $ dN_{\omega}/(d{\omega}d{\mu}) \sim 
 {\omega}^{-\alpha} {\rm exp} (-{\omega}/kT_{c})$,
 where $\omega$ is the photon energy, $\alpha$ is the
 power law index and $T_{c}$ the continuum temperature. 
 Due to the steepness of the input spectrum, the photons are injected
 with a flat spectral distribution, and are assigned appropriate weight factors.
 This improves the statistics, but upscattered photons with large weight
 factors create some noise level in the high energy part of the spectrum.
 The line parameters are: the electron temperature $T_e$, the local 
 magnetic field $B$ (assumed uniform) and the minimum value of continuum
 optical depth $\tau_{c}$ (in the direction parallel to $B$ for a slab
 and perpendicular to $B$ for a cylinder). 

    A discussion on the variation in the line shapes with different 
 combination of parameters is given below but first we draw attention
 to the effect of multiple scatterings on the line features and thus,
 on our selection of parameters. Multiple scattering occurs              
 mostly for first harmonic photons; higher harmonic photons are 
 more likely to undergo single resonant Raman scattering, leaving the electron 
 in an excited Landau level. Subsequent decay of these states produces
 spawned photons that, in essence, amounts to additional photon injection
 shortward of the fundamental energy (i.e. due to the anharmonic spacing of
 relativistic resonant energies). As $B$ approaches $B_{\rm cr}$ 
 multiple scattering becomes increasingly more important for
 higher harmonics.  
 The line profiles are narrower and deeper                                  
 at large angles and broader and shallower at small angles due in part
 to one-dimensional Doppler broadening. Thus, in a cylindrical scattering
 region, the lines are deepest at small angles, where the optical depth 
 is largest: $\tau^{cy}_c(\theta) = \tau_c / {\rm sin}\theta$;
 while in the slab geometry the lines are deepest at large angles,
 where outgoing photons encounter the largest optical depth:
 $\tau^{sl}_c(\theta) = \tau_c / {\rm cos}\theta$ (compare Figs 2d and 2a).
 Thus, spawned photons and angle redistribution 
 of photons through multiple scattering produce complex structure in 
 the emerging fundamental line
 features.  This influences the determination of model parameters in 
 several ways.  First, the inferred (from observation) hardness of     
 the continuum may be strongly affected by broad line wings.  Ideally, the
 fitting of three `uncontaminated' continuum flux points would determine
 $\alpha~ {\rm and}~ T_c$. However, for the particular case of the OSSE
 observation, choosing the third flux point on the blue side of the 
 110 keV line may violate this condition (see Fig. 2e). 
  Second, although the resonances occur at well defined energies
 (${\omega}_{res}^n/m_e = 
 [(1 + 2~n~b~{\rm sin}^2\theta)^{1/2} - 1] / {\rm sin}^2 \theta$;
 where $b \equiv B/B_{\rm cr}$), the structure in the line introduces a
 larger uncertainty in the evaluation of the field strength than does
 the assumed viewing angle. 

  Third, the electron temperature cannot be easily estimated from the line
 widths.  An estimate of the equilibrium electron temperature for A0535+26
 can be found from the expression $T_e = \omega_{\rm cyc}/(2+\alpha_{eff})$
 of Lamb, Wang and Wasserman (1990), which is valid in the limit of a single 
 scattering.  In this approximation, the effective spectral indices of the 
 continuum at the fundamental, $\alpha_{eff} \sim 6.5$ and 3.6, yield
 $T_e \simeq 12.9$ keV and 9.8 keV for the high and low field model electron
 temperatures respectively .  However, in our modeling 
 $T_{e}$ must be considered a free parameter determined from
 the `symmetry' of the resulting line shapes.  Empirically we observe that too
 large a temperature results in blue excess emission mostly at small angles
 (Fig: 2c and 2f); on the other hand if $T_{e}$ is too low, red excess emission 
 is produced (Fig: 2a and 2d). 
 
   \begin{figure*}
\epsfysize=12cm
\epsfbox{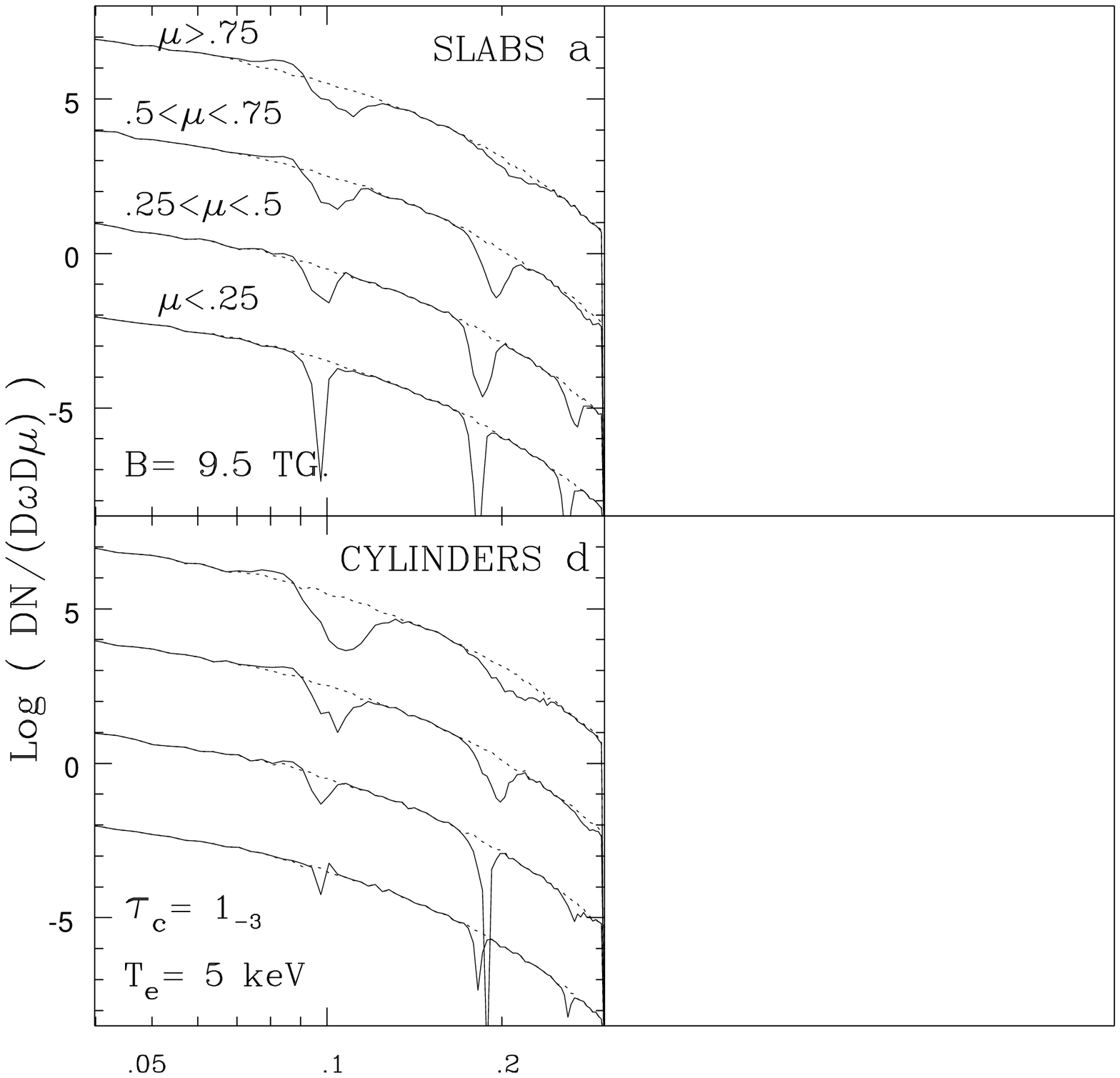}
\vskip -12.035cm
\hskip 5.03cm
\epsfysize=12cm
\epsfbox{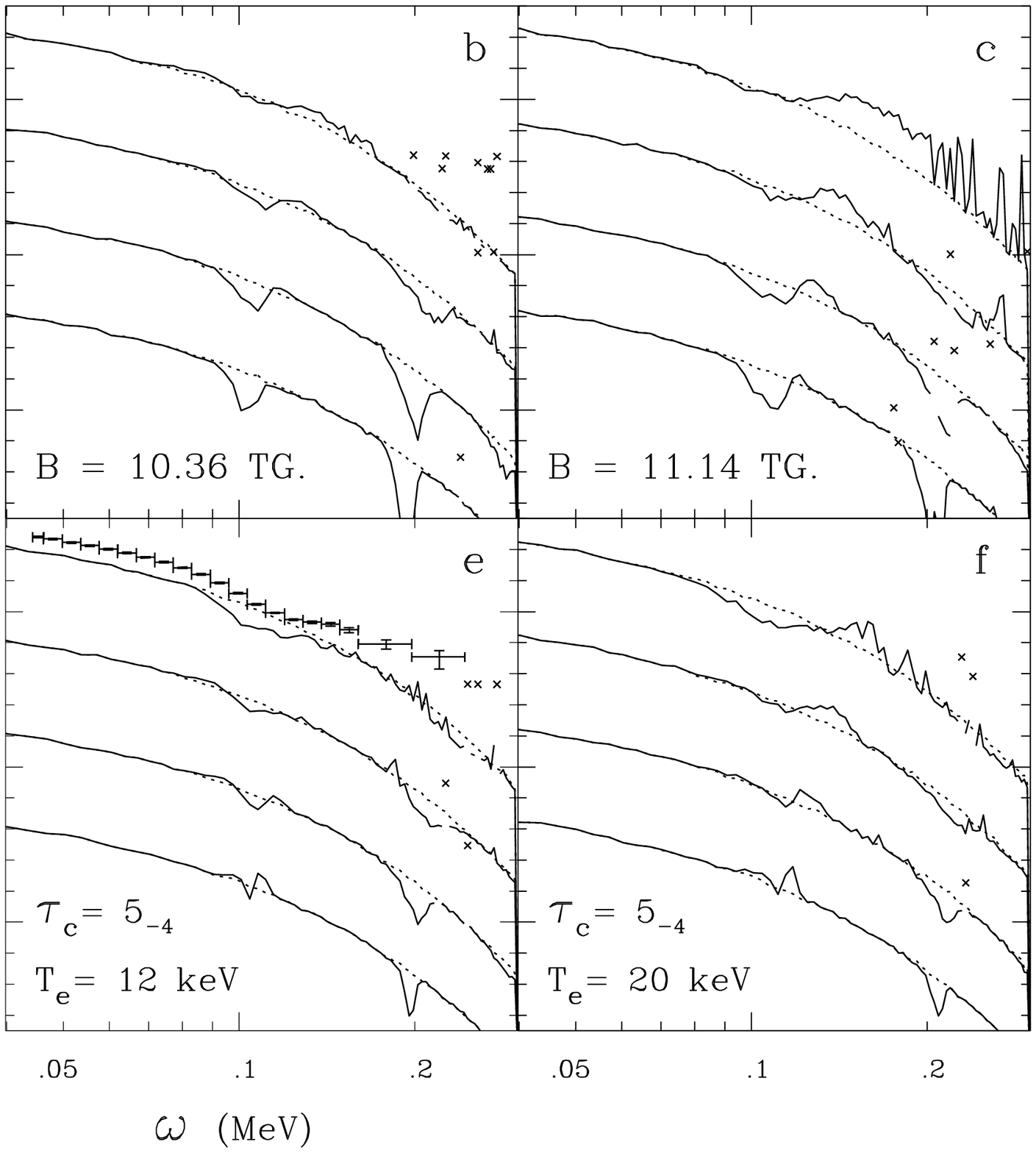}    
\caption{
 Angle dependent model photon spectra for higher field cases.
   $T_c=17.8$ keV and $\alpha = -.13$ for the two leftmost spectra 
   and 14.5 keV and 0 for the rest. Error bars are the A0535+26 data.
   }
               
         \label{Fig2}%
    \end{figure*}
\vskip 2.0cm

\subsection{The Lower Field Case} 

    Lower field models, shown in Fig 1, have $B/B_{\rm cr} \sim .12$, 
 $\omega_{res}(\mu=0) = 57.5$ keV, and are constrained here so 
 that either very narrow or very 
 shallow first harmonics result (i.e. small equivalent width).
 Although spawned photons may produce additional injection around the 
 fundamental energy, the steepness of the continuum in A0535+26 makes
 this process ineffective.  For example, in the model shown in Fig 1c,
 spawned photons contribute only $\sim 1\%$ of the flux at the first
 harmonic for $.5<\mu<.75$).
 Nevertheless, `filling in' at the fundamental energy may be attained in
 a cylindrical geometry, at a large viewing angle, through scattered photons
 from smaller angles (Figs 1c and 1d). Photons trapped close to the line
 center for $\mu \sim 1$ escape more easily when scattered into
 $\mu \sim 0$ where the line profiles are narrowest. However, 
 for $\mu \simeq 0$, this results in measurable $emission$ at the 
 fundamental energy. Moreover, the resulting line widths at 110 keV are
 too narrow when compared to the data and increasing  $T_e$ (Fig. 1c) does 
 not widen the second harmonic noticeably. For a slab,
 the effect of an overly high $T_{e}$ is much more pronounced (Fig. 1d).
    Note that both geometries yield a very shallow fundamental feature
 for the angle bin $.5 < {\mu} <.75$ (Figs. 1a and 1c) and a wide
 second harmonic feature for a reasonable temperature: $T_{e} \sim$ 12 keV.
 Nevertheless, the feature at the fundamental energy is at least 
 half the depth of the second harmonic, and would have been clearly
 identifiable by the OSSE observations (shown with error bars in Fig 1c)
 as it spans about two energy bins of the OSSE instrument.

\subsection{The Higher Field Case} 
    
 In our high field models, $B/B_{\rm cr} \sim .21-.25$
 and $\omega_{res}^1(\mu=1)=120-129$ keV. We find the parameters of 
 the scattering
 region which are best able to produce a first harmonic feature similar to
 the observed line at 110 keV. When account is made for the structure in
 the line, fields as high as 11 TG may be consistent with the observed line.
 Fig 2 exhibits a sample of high field model spectra, and illustrates that 
 the closest line feature to that of A0535+26 is provided by
 a cylindrical geometry (Fig. 2e) and for angle bin $.75 < \mu < 1$. 
 The model has three free parameters: B, $\tau_c$ and $T_e$.

\section{Conclusions}

 Based on our theoretical models of the spectrum of A0535+26, assuming
 both low and high magnetic fields and a range of electron temperature,
 optical depth and continuum spectrum shape, we conclude that the observed 
 110 keV feature is more likely to be a first harmonic.  Our conclusion is,
 however, limited by the fact that we have thus far made only qualitatively
 comparisons between our model spectra and the OSSE photon spectrum.
 The latter has already assumed a particular line and continuum shape 
 to unfold the observed count spectrum.  Given that escape of cyclotron
 photons into the line wings can significantly affect the surrounding
 continuum, it is important to fit both the line and continuum spectrum
 simultaneously. We are planning to carry out formal fits to both the
 OSSE and HEXE/TTM phase-resolved spectra, folding our model spectra through
 the response functions of the detectors.  It will then be possible to make
 a more quantitative
 statement about the likelihood of a high field ($B \sim 10$ TG) vs. a 
 low field ($B \sim 5$ TG) in A0535+26.

 \begin{acknowledgements}

 We thank Dr. Ramin Sina for aid in the scattering cross section code
 and Dr. Alexander Szalay for allowing us access to speedy computer resources
 at the Johns Hopkins University.
      
\end{acknowledgements}

\end{document}